\documentclass[12pt,preprint]{aastex}
\usepackage{graphicx}
\usepackage{natbib}
\usepackage[usenames,dvips]{color}
\begin{document}

\title{Spitzer Space Telescope Observations of the Magnetic 
Cataclysmic Variable AE Aqr}

\author{Guillaume Dubus}
\affil{Laboratoire d'Astrophysique de Grenoble, UMR 5571 CNRS Universit\'e Joseph 
Fourier, BP 53, F-38041 Grenoble, France}
\email{gdubus@obs.ujf-grenoble.fr}

\author{Ronald E. Taam}
\affil{Northwestern University, Department of Physics and Astronomy,
  2131 Tech Drive, Evanston, IL 60208; ASIAA/National Tsing Hua University -
 TIARA, Hsinchu, Taiwan}
\email{r-taam@northwestern.edu}

\author{Chat Hull} 
\affil{University of Rochester, Department of Physics and Astronomy, 
  Rochester, NY 14627}
\email{chat.hull@gmail.com}

\author{Dan M. Watson} 
\affil{University of Rochester, Department of Physics and Astronomy, 
  Rochester, NY 14627}
\email{dmw@pas.rochester.edu}

\and

\author{Jon C. Mauerhan} 
\affil{University of California, Department of Physics and Astronomy,
  Los Angeles, CA 90095} 
\email{mauerhan@astro.ucla.edu}

\begin{abstract}
The magnetic cataclysmic variable AE Aquarii hosts a rapidly
rotating white dwarf which is thought to expel most of the material
streaming onto it.  Observations of AE Aqr have been obtained in the
wavelength range of $5 - 70 \mu$m with the IRS, IRAC, and MIPS
instruments on board the {\it Spitzer Space Telescope}.  The
spectral energy distribution reveals a significant excess above the
K4V spectrum of the donor star with the flux increasing with
wavelength above 12.5 $\mu$m. Superposed on the energy distribution
are several hydrogen emission lines, identified as Pf $\alpha$ and
Hu $\alpha$, $\beta$, $\gamma$.  The infrared spectrum above 12.5
$\mu$m can be interpreted as synchrotron emission from
electrons accelerated to a power-law distribution $dN\propto
E^{-2.4}dE$ in expanding clouds with an initial evolution timescale 
in seconds.  However, too many components must then be superposed to 
explain satisfactorily both the mid-infrared continuum and the 
observed radio variability. Thermal emission from
cold circumbinary material can contribute, but it requires a disk
temperature profile intermediate between that produced by local
viscous dissipation in the disk and that characteristic of a
passively irradiated disk.  Future high-time resolution observations
spanning the optical to radio regime could shed light on the
acceleration process and the subsequent particle evolution.
\end{abstract}


\section{Introduction}

Multi-wavelength observational investigations of cataclysmic variable
(CV) binaries are of continuing interest since they provide important
diagnostic information constraining the characteristics of the white
dwarf, main sequence like donor, and accretion disk in these systems.
Within the last decade, observational studies of CVs at near to mid
infrared wavelengths have been carried out in order to determine the
spectral type of the mass losing star.  Further interest in long
wavelength studies of CVs stems from the possibility of detecting cool
gas surrounding such systems as its presence can have some affect on the 
secular evolution of these systems (Spruit \& Taam 2001).
Circumstantial evidence for the presence of this 
material had been provided by the existence of faint features characterized 
by narrow line widths in the optical and far UV spectrum of several objects 
(e.g., see references in Dubus et al. 2004), although such features are not 
present in all systems (Belle et al. 2004).  Recently, {\it Spitzer Space 
Telescope} observations have revealed the presence of excess infrared emission 
in magnetic CVs (Howell et al.  2006; Brinkworth et al. 2007) and black hole 
low-mass X-ray binaries (Muno \& Mauerhan 2006).
 
In view of the accumulating observational evidence for the presence of
cool gas surrounding CVs and its possible importance for CV evolution, 
AE Aquarii was observed with the {\it Spitzer Space Telescope}.  \object{AE 
Aqr} is a particularly unusual CV as the white dwarf in the system rotates (at
33.08 s) asynchronously with the orbital motion ($P_{orb} = 9.88$
hrs).  It is highly variable exhibiting flaring behavior in the
optical (van Paradijs et al. 1989), radio (Bastian et al. 1988), and
ultraviolet (Eracleous \& Horne 1996) wavelength regions.  More recent
observations of AE Aqr in the mid infrared and far infrared wavelengths 
using ISO (Abada-Simon et al. 2005) have led to the measurement of a
crude spectral energy distribution (SED) from $3.6 - 170 \mu$m.
Observations in the mid infrared wavelength region were also carried
out by Dubus et al. (2004) at {\it Keck Observatory} leading to the
detection of excess emission in AE Aqr at 12 $\mu$m above that
extrapolated from shorter wavelengths and variable on timescales less
than an hour.

To significantly improve on the measurement of the SED in the
wavelength range from $5 - 70 \mu$m, to identify any possible thermal
contribution and to further test the synchrotron emission
interpretation for the mid-infrared, based on the multiple injection
expanding clouds model (Bastian, Dulk, \& Chanmugam 1988) successfully
used in the radio regime, we report on the observations of AE Aqr
based on measurements obtained with the IRAC, IRS (Infrared
Spectrometer), and MIPS (Multiband Imaging Photometer) instruments on
the {\it Spitzer Space Telescope}. Recently, Harrison et al. (2007) 
independently reported on the IRS observations.  In the next section, we describe
the observations and the method of analysis of the {\it Spitzer}
infrared data.  The SED of AE Aqr from combined IRAC, MIPS, IRS and
archival data is presented in \S3.  The implications of these results
for CVs are discussed in \S4 and summarized in the final section.
 
\section{Observations}

\subsection{IRAC/MIPS}

Observations of AE Aqr were performed with the Infrared Array Camera (IRAC; Fazio et 
al. 2004) in all four channels (3.6, 4.5, 5.8, and 8.0 $\micron$ ) beginning Julian 
date 2453501.1. Images were obtained in subarray mode, producing 2 composites of 64 
$\times$ 0.1 sec exposures in a 4 point dither pattern for each channel. The final mosaic 
images were produced for photometric extraction using the Basic Calibrated Data products 
(pipeline version S12.0.2). Aperture photometry was performed with the IRAF APPHOT package 
using the standard IRAC 10 pixel aperture radius (12$\arcsec$.2) with a sky annulus 
extending 10 to 18 pixels (IRAC Data Handbook, Version 3.0, 2006). Since a high SNR 
($>$ 23-167) was achieved in all four IRAC channels, the photometric errors in these 
bands are dominated by an absolute calibration uncertainty of 10\%, resulting from the 
not-yet fully characterized IRAC filter bandpass responses in subarray mode (Quijada et 
al. 2004, Reach et al. 2005, Hines et al. 2006). The results are listed in Table 1. 

AE Aqr was also imaged with the Multiband Imaging Photometer for Spitzer (MIPS; Rieke et 
al. 2004) at 24 $\micron$ and 70 $\micron$ beginnning Julian date 2453503.8.  A single 
3 sec exposure was obtained at 24 $\micron$ and a 10 $\times$ 10 sec composite was obtained 
at 70 $\micron$. Photometry was extracted from the Post Basic Calibrated Data products 
(pipeline version S12.4.0) using the IRAF APPHOT package. A 13$\arcsec$ aperture radius 
with a sky annulus extending 20$ \arcsec$ to 32$\arcsec$ and aperture correction of 1.167 
was used for the MIPS 24 $\micron$ image, whereas, a 35$\arcsec$ aperture radius with sky 
annulus extending 39$\arcsec$ to 65$\arcsec$ and aperture correction of 1.21 was used for 
the 70 $\micron$ image. The SNR at 24 $\micron$ and 70 $\micron$ was sufficiently high 
($>$100 and 10, respectively) that the photometric uncertainty is dominated by absolute 
calibration errors of 10\% and 20\% for MIPS 24 $\micron$ and 70 $\micron$, respectively 
(MIPS data handbook, Version 3,2, 2006). The total unceratinty for the 70 $\micron$ 
measurement was obtained by summing in quadrature the 20\% absolute calibration error 
with the inverse of the SNR, resulting in a final error of 22\%. The results are also 
listed in Table 1.

\subsection{IRS}

IRS\footnote{The IRS was a collaborative venture between Cornell University and Ball 
Aerospace Corporation funded by NASA through the Jet Propulsion Laboratory and the 
Ames Research Center.} observations of AE Aqr were carried out on 2004 November 13
(Spitzer AOR ID 12699904), as part of the IRS\_Disks guaranteed-time
observing program.  In somewhat different form, the data were analyzed
and discussed by Harrison et al. (2007).  We used the IRS long-slit,
low-spectral-resolution modules ($\lambda /\Delta\lambda ~=~60-100$)
to record the 5.3-36 $\mu$m spectrum of AE Aqr. Four 14-second
exposures were taken at $\lambda~=~5.3-14~\mu$m, and six 30-second
exposures were taken at at $\lambda~=~14-36~\mu$m, divided equally in
each case between two sets of observations with the object nodded
along the slit by a third of a slit length. We reduced the resulting
spectra using the SMART software package (Higdon et al. 2004).  Starting with
non-flatfielded ("droop") 2-D spectral data products from the Spitzer
Science Center IRS pipeline (version S14), we removed sky emission
by subtraction of differently- nodded observations, and extracted
signal for each within a narrow window matched to the instrumental
point-spread function.  The resulting 1-D spectra were multiplied by
the ratio of a template spectrum of the A0V star $\alpha$ Lac (Cohen et al. 2003), 
to identically-prepared IRS observations of this star, to produce
calibrated spectra for each exposure.  Averaging all of the exposures,
we obtain the final spectrum shown in Figure 1. 

Like most grating spectrographs, the modules of the IRS respond better
to light polarized parallel to the grating rulings than to the
orthogonal linear polarization.  The ratio of the response in the two
linear polarizations is, however, poorly characterized at present.
Our observations of AE Aqr were done with the $\lambda~=~5.3-14~\mu$m
slit at position angle -22.5\degr , and the $\lambda~=~14-36~\mu$m
slit nearly perpendicular to the other, at -106.1\degr.  These slit
orientations are fixed with respect to each other, and cannot be
adjusted significantly on the sky for objects as close to the ecliptic
plane as AE Aqr.  It appears from the continuity of the IRS spectrum at
14 $\mu$m (difference $<4\%$ of the signal) that AE Aqr does not have
a large linear polarization in either slit direction at this
wavelength. It is possible that the object is more highly polarized
at the longest IRS wavelengths than it is at 14 $\mu$m, and
because we cannot check the orthogonal polarization at long wavelengths, this
possibility adds an unknown additional uncertainty to the
flux calibration there; at shorter wavelengths we estimate the
photometric uncertainty to be 5\% (1$\sigma$).

\section{Results}

The SED based on the fluxes obtained from the IRS instrument is
illustrated in Fig. 1. It can be seen that the flux in the spectrum
exhibits a significant excess at longer wavelengths compared to 
the expected Rayleigh Jeans contribution of the K4V companion star.  The 
flux actually increases at wavelengths longer than 12.5 $\mu$m, as noted by
Harrison et al. (2007). In addition to the hydrogen emission lines at 7.5 $\mu$m
(Pf $\alpha$ and Hu $\beta$), and 12.4 $\mu$m (Hu $\alpha$) identified
by Harrison et al. (2007), we also find evidence for the Hu $\gamma$ transition at
5.9 $\mu$m. 

The sensitivity of the observations was sufficient to investigate
variability of AE Aqr on the scale of the times of individual
exposures.  Figure 2 is a plot of the flux in the Pf $\alpha$/Hu
$\beta$ blend ($\lambda~=~7.5~\mu$m), and of the continuum flux within
bands at $\lambda~=~20-28~\mu$m and $28-32~\mu$m, as functions of
time. Here, the uncertainties in Fig. 2 were propagated from the 
IRS noise based uncertainties in each wavelength channel of the individual
exposures and (in the case of Pf $\alpha$) the uncertainty in the fit of 
a Gaussian profile to the line. Although there is a hint of substantial 
variability in Pf $\alpha$/Hu $\beta$ on minute time scales, the statistical
significance is not high (about 2$\sigma$), and thus remains to be
confirmed with a longer observation.  The long-wavelength continuum
bands show no significant variation on minute timescales. Given
the uncertainties in each measurement, detecting variability at the
3$\sigma$ level would have required a deviation of 50-100\% from the
mean flux in one bin. In comparison, Dubus et al. (2004) found 30-50\%
variability between flux measurements taken about an hour apart at
4.6, 11.3 and 17.6 $\mu$m (with integration times of several
minutes).

The overall SED including the fluxes obtained by the IRAC and MIPS
instruments is shown in Fig. 3. The fluxes obtained from ISO and IRAS
measurements (Abada-Simon et al. 2005), average radio and mm fluxes
from Abada-Simon et al.  (1993), and {\it Keck} measurements in the
optical and near infrared are also plotted (Dubus et al. 2004) to
provide a spectrum extending over 6 orders of magnitude in frequency.
For comparison, the spectrum of a K4V star is fitted to the optical 
and near infrared data (see Dubus et al. 2004 for details) and 
is plotted overlaying the SED to illustrate the presence
of excess infrared emission for wavelengths longer than 5.8 $\mu$m.
The flux measurements for the two long wavebands of the IRAC
instrument are not only consistent with those obtained from the IRS
instrument, but also with the {\it Keck} mid-IR fluxes obtained in
2002.  Overall, the average infrared spectrum from 10-100 $\mu$m is well
approximated by a $\nu^{-0.7}$ power-law.

\section{Discussion}

\subsection{Non-thermal synchrotron radiation\label{nonthermal}}
Non-thermal synchrotron radiation provides a common framework to
interpret the SED from radio to mid-infrared frequencies. 
Although Harrison et al. (2007) considered cyclotron emission from 
the white dwarf, they concluded that the synchrotron radiation 
interpretation was more likely, confirming earlier work of Dubus 
et al. (2004). The inverted radio spectrum is typical of superposed
emission from multiple synchrotron self-absorbed components, as seen
in X-ray binary and AGN jets. Indeed, Bastian, Dulk, \& Chanmugan 
(1988) modelled the radio
emission by multiple clouds of particles undergoing adiabatic
expansion. The average spectrum then steepens around $2\times
10^{12}$~Hz to $F_\nu\propto \nu^{-0.7}$, as expected for optically
thin synchrotron emission from electrons in freshly injected clouds.
The required index $p=2.4$, assuming a power-law distribution
$dN\propto E^{-p}dE$, is the canonical index derived from shock
acceleration theory.

As described in Bastian et al. (1988), the
clouds are characterized by their initial radius $R_0$, particle
density $n_0$ and magnetic field $B_0$. These values probably change
from cloud to cloud but are assumed identical here (representing a
time-averaged value).  The expansion scales with $\rho\equiv
R/R_0=(1+t/\beta t_0)^\beta$ with $t_0 \equiv R_0/v_0$, and $v_0$ the
initial expansion velocity. Reproducing the average radio slope
$\nu^{0.5}$ with $p=2.4$ requires $\beta\approx 0.53$ (see Eq.~1 in
Dubus et al. 2004), faster than expansion in a uniform medium ($\beta$=2/5)
but slower than steady expansion ($\beta$=1). Adiabatic cooling
arguably dominates since the infrared spectrum up to a few
$10^{13}$~Hz appears unaffected by synchrotron or inverse Compton
losses, which would steepen the power-law at high frequencies.
Requiring that the timescale for adiabatic losses ($t_0$) be shorter
than the synchrotron timescale of electrons emitting at $10^{13}$~Hz
sets an upper limit to $t_0$ of $2.4\times 10^5 B_0^{-3/2}
\nu_{13}^{-0.5}$~s.

Although the average SED can be reproduced by multiple clouds (see
below), reconciling this interpretation with the variability leads to
two puzzles. The first is that the variability is not as
strong as expected if a single cloud dominated the infrared SED. In
this case, the initial optically thin emission would decrease rapidly
with time ($F_\nu\propto \rho^{-2\beta p}$). Either most of the
variability occurs on longer timescales than those sampled in
individual observations ($t_0\ga 1000$ s) or it occurs very quickly
($t_0 \la $ 30 s, Fig.~2) and is smoothed out by the continuous
ejection of new clouds combined with a coarse time resolution. A long
$t_0$ is unlikely as radio observations would then show little
variability: the peak frequency for a single cloud moves from
far-infrared to radio frequencies on a timescale $t\sim 100\, t_0
(\nu_{12}/\nu_9)^{0.77}$ so that the variation in radio would be
extremely slow. Inversely, observations of radio variability on ks
timescales directly set $t_0\approx 10$~s. The infrared emission is
then the average flux from clouds evolving on a sub-min timescale with
an average elapsed time between cloud ejection $t_{\rm flare}\la t_0$.
This flaring timescale cannot be too long compared to $t_0$ (or the
peak flux would need to be very high to average out to $\approx
100$~mJy at $2\times 10^{12}$~Hz), nor can it be too short (since then
the average SED would require superposing many clouds with small peak
flux, implying little variability at any wavelength). The most
plausible case appears to be $t_{\rm flare}\sim t_0\sim 10$~s. In
this case, each 30s bin in Fig.~2 would be an average of three
consecutive flares, explaining why the lightcurve shows
little variation over 6 min.

Figure~3 shows the average SED expected using $t_{\rm
flare}=t_0=10$~s, $n_0=4.7\times 10^{10}$~cm$^{-3}$, $R_0=1.3\times
10^9$~cm and $B_0=1830$~G. To obtain these numbers, which are comparable 
to those found from the analysis of the radio data by Bastian et al. (1988), 
an equipartition magnetic field
was assumed and the values of the initial peak emission (210~mJy at
$3\times 10^{12}$~Hz) were adjusted so as to have the average spectrum
peak at about 100 mJy at $2\times 10^{12}$~Hz.
The average flux corresponds to the
emission from a single cloud integrated over time and multiplied by
the flaring rate $t_{\rm flare}^{-1}$. 
However, if the peak is {\em on average} at 40 $\mu$m with a flux 
of $\approx 50$~mJy, then the multiple clouds interpretation will 
fail to reproduce the average radio flux.

With the peak position fixed, $B_0$, $R_0$ and $n_0$ are set uniquely
by the observations in as much as equipartition is assumed and $t_{\rm
  flare}=t_0$ (regardless of the actual value). A weaker (sub-equipartition) magnetic
field would be actually preferable as the synchrotron timescale at
10$^{13}$~Hz for the equipartition field (3~s) is smaller than the
adiabatic timescale (10~s). The cloud size compares well with the size
of the magnetized blobs through which mass transfer has been proposed
to occur in AE Aqr. These blobs approach the white dwarf down to
$\approx 10^{10}$~cm at which point the spinning magnetic field of the
white dwarf presumably propels them out of the system. Note that
$P_{\rm spin}\sim t_0$ and that the escape velocity at closest
approach is comparable to the expansion velocity $v_0=R_0/t_0\approx
1300$~km~s$^{-1}$.

Figure~3 also shows the instantaneous emission from a cloud at $t=0$
and $t=t_0$ (before a new ejection occurs). Strong variability is
expected on a timescale of a few seconds (less than that sampled
here by {\it Spitzer}) and this can be tested by a high-time
resolution mid or far-IR light curve. Emission to near-IR frequencies
require electrons of a few 100 MeV. The maximum electron energy is
arbitrarily set at 500~MeV in Fig.~3. The high energy electrons
contribute a little to the optical $V$-band flux, but this is not
sufficient to explain the 0.5~mag flaring that is observed on
timescales of minutes (van Paradijs, Kraakman \& van Amerongen 1989).  
Although the flaring at optical
(and X-ray) frequencies has also been attributed to propelled gas
(Eracleous \& Horne 1996), the connection with the non-thermal flaring at low
frequencies remains obscure.

The rapid cloud evolution illustrated in Fig.~3 raises the second
puzzle. The peak flux from a single cloud varies as $S_{\rm
  peak}\propto \nu_{\rm peak}^{-1.3}$ so that a cloud initially
emitting 100~mJy in infrared can only emit a few 0.1~mJy at most in
radio. The radio emission therefore consists of the superposition of
hundreds of faint clouds, at odds with the observations of variation
of 1-10~mJy in amplitude. A possible solution is that the clouds are
re-energized during their expansion (Meintjes \& Venter 2003). 
The variability properties would change,
enabling longer variations in infrared to be compatible with the radio
flares. Again, a comparison between lightcurves at IR and radio
frequencies, notably to characterize lags, would shed light on the
conditions during expansion.

\subsection{A thermal component? \label{thermal}}
The circumstantial evidence points towards non-thermal emission.
However, in the absence of variability directly linking the infrared
emission to the radio, we cannot exclude a contribution from circumbinary  
material. A multicolor disk blackbody fit to the infrared requires a
temperature distribution $T\propto R^{-0.54}$, in between the $-3/7$
slope of a thin disk passively heated by irradiation and the
$-3/4$ slope of a thin disk heated by viscous dissipation.
We note that a profile of a similar form, $T\propto R^{-1/2}$,
has been found from detailed vertical structure models of irradiated
accretion disks (D'Alessio et al. 1998) to provide satisfactory
fits to the emission properties of young stellar objects (D'Allesio
et al. 1999).

The {\em Spitzer} data is fitted adequately (Fig.~3) by such a disk
extending out to 1.2~AU at a temperature of $55$~K, taking a distance
of 102 pc (Friedjung 1997) and an inclination of 55$\degr$. 
Here, the disk peaks around 40 $\mu$m. This is a better fit than a 
single temperature (140~K) blackbody (Harrison et al. 2007). Optically 
thin emission from material at
larger distances, and not taken into account here, may contribute to
longer wavelengths. The dominant contribution below $\sim 10^{12}$~Hz
would still be non-thermal flaring. The contraints on the flare peak
frequency and flux are therefore relaxed compared to
\S\ref{nonthermal}, but not enough to account for the amplitude of the
radio flares without {\em e.g.} re-energizing the clouds.

The expected infrared variability from CB material is slow since the
thermal timescale is roughly Keplerian hence $\ga P_{\rm orb}$.
Variability on timescales of years might explain the discrepancy
between the {\em Spitzer} and ISO far-infrared measurements. However,
the disk would have to be colder and larger to account for the ISO
flux at 90 $\mu$m.  The variability seen on a sub-hour timescale by 
Dubus et al. (2004) argues against thermal infrared emission but has 
yet to be confirmed by a more extensive set of observations. 
A disk could be resolved by
interferometric observations at mm wavelengths, or possibly in mid-IR
where the emission is already a few milli-arcseconds wide.
Polarimetric observations may also distinguish between scattered light
and synchrotron emission.

\subsection{Infrared line emission}
The detection of Pf $\alpha$ and Hu $\alpha$, $\beta$ and $\gamma$ in
the IRS spectrum is consistent with observations of intense Brackett
and Pfund lines in the near-infrared from AE Aqr (Dhillon \& Marsh 1995). 
These lines were attributed to the accretion disk due to their intensity and
their widths of $\approx 1500$~km\, s$^{-1}$ (unfortunately below the
IRS spectral resolution). However, high temporal and spectral
resolution studies of H$\alpha$ show no disk in AE Aqr and that most
of the mass flow from the donor star is probably ejected by the
spinning magnetosphere before reaching the white dwarf (Wynn et al. 1997;
Welsh et al. 1998).
The hydrogen lines are probably produced in the propeller outflow. We
note that the combination of intense H line emission and radio
emission is reminiscent of outflows in young stellar objects -- with
the distinction that the radio emission is due to bremsstrahlung in
the latter ({\it e.g.} Simon et al. 1983).

\section{Conclusion}
The presence of infrared emission in excess of expectations from the
stellar companion in AE~Aqr, already present in Keck data up 
to 17 $\micron$ (Dubus et al. 2004) and in ISO data (Abada-Simon et al. 2005), 
is confirmed by {\it Spitzer}. Synchrotron
emission from multiple expanding clouds provides a coherent framework
to interpret the SED from radio to infrared wavelengths.
Alternatively, interpreting the infrared SED as thermal emission
from CB material provides an intriguing parallel with disks 
surrounding T Tauri stars. This interpretation would still 
require non-thermal flaring to
explain the radio emission and would be ruled out if fast, large
amplitude variability is confirmed at infrared wavelengths.  On the
other hand, synchrotron emission from electrons injected with a
canonical $E^{-2.4}$ power-law reproduces well the whole spectrum
down to radio frequencies.

However, the multiple cloud picture raises several conundrums.  Rapid
(seconds) variability is expected in the infrared, but remains
undetected, probably for lack of an adequately sampled light curve.
In addition, the number of clouds required to reproduce the average
radio spectrum is too large to also explain the amplitude of the
variations seen at these frequencies (Meintjes \& Venter 2003). This requires
modification of the basic assumptions adopted by Bastian, Dulk, \& 
Chanmugam (1988). A
continuous outflow from the propeller, rather than discrete ejections,
may provide a better description of the process operating in AE Aqr.
Other questions remain open, such as the nature of the particle
acceleration process and the link to the flaring behaviour at optical
to X-ray frequencies.  Better multiwavelength sampling for short
time scale ($< 30$ s) spectral evolution studies is needed to obtain further 
insights into the physics of this unique system.

\acknowledgments We are grateful to Michael Jura for several helpful 
discussions.  This work was supported, in part, by the Theoretical
Institute for Advanced Research in Astrophysics (TIARA) operated under
Academia Sinica and the National Science Council Excellence Projects
program in Taiwan administered through grant number NSC 95-2752-M-007-006-PAE, 
by NASA through the IRS Instrument Team (JPL contract 1257124), and by 
the NSF Research Experience for Undergraduates program at the University of Rochester. 

{\it Facilities:} \facility{Spitzer (MIPS, IRAC, IRS)}

\begin{table}
\caption{{\it Spitzer} Photometry of AE Aquarii}
%\end{center}
%\begin{center}
\begin{tabular}{cl}
\hline
\hline
Wavelength &  Flux \\
(${\mu}$m )& (mJy) \\
\hline
3.6 & 138.9 ${\pm}$ 13.9  \\
4.5 &  73.2 ${\pm}$ 7.3 \\
5.8 &  56.1 ${\pm}$ 5.6 \\
8.0 & 38.2 ${\pm}$ 3.8 \\ 
24 & 39.2 ${\pm}$ 3.9  \\
70 & 52.5 ${\pm}$ 11.6 \\
\hline
\end{tabular}
\end{table}
\clearpage

\begin{figure}
\center
\resizebox{\hsize}{!}{\includegraphics{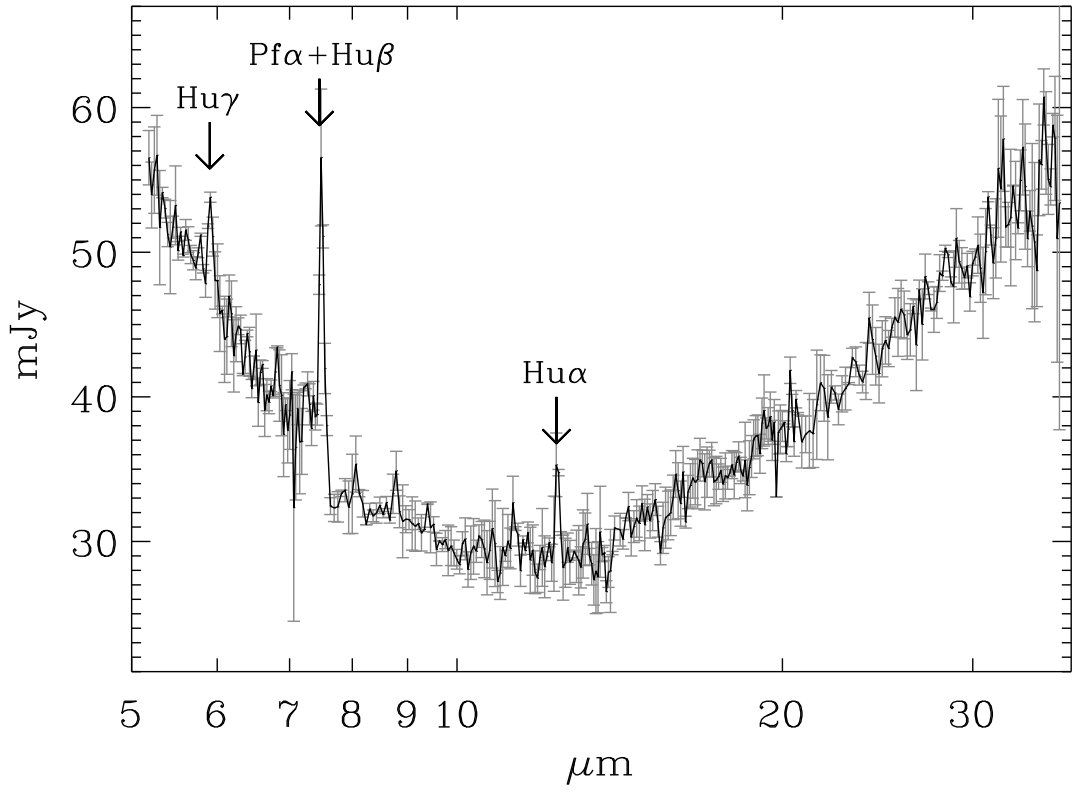}}
\caption{IRS SL-LL spectrum of AE Aqr with line detections indicated.
The uncertainties in the fluxes, as derived from the standard 
deviation from the mean of individual exposures are also shown.}
\end{figure}

\begin{figure}
\center
\resizebox{\hsize}{!}{\includegraphics{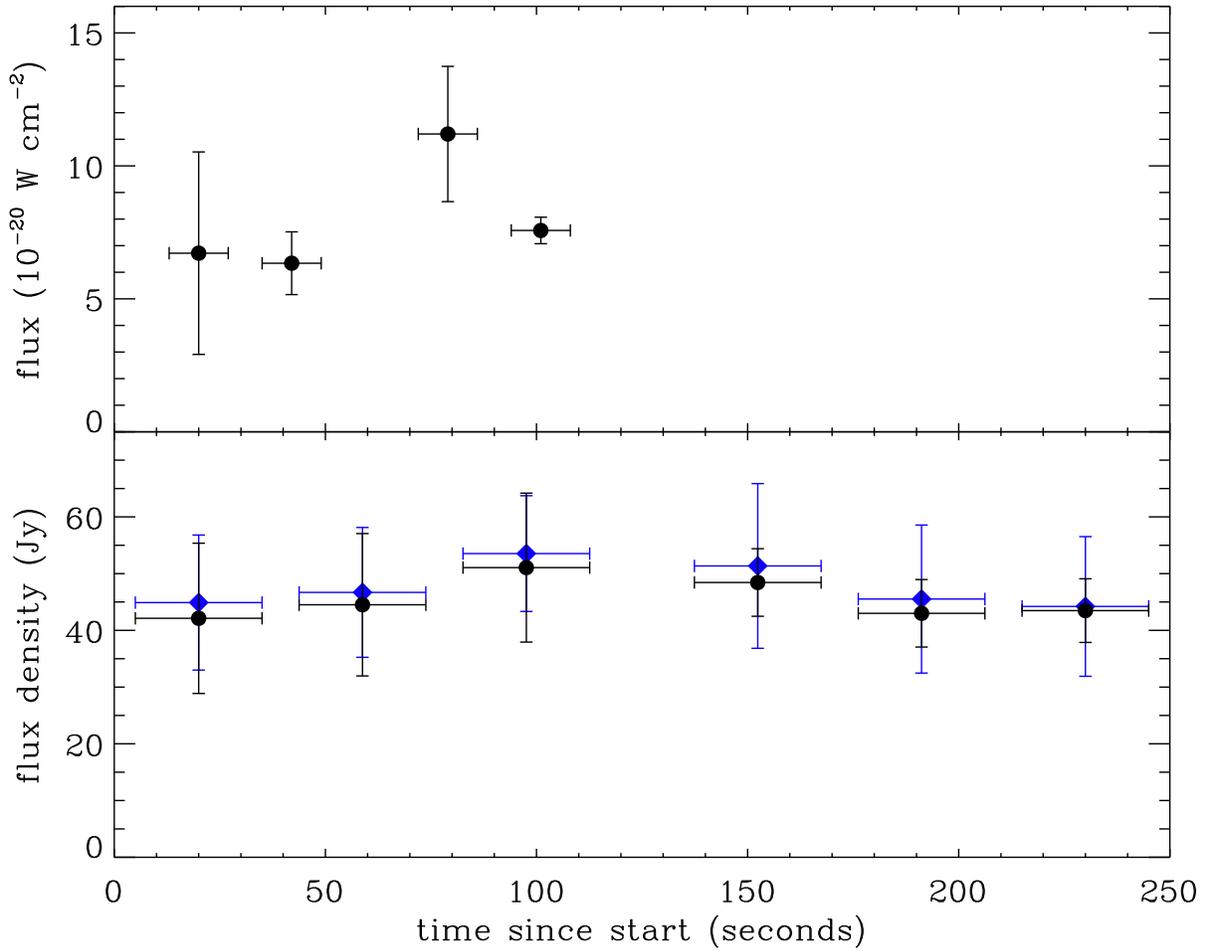}}
\caption{Time dependence of signals during observations. Top panel:
flux in the Pf $\alpha$/Hu $\beta$ line blend. Bottom: flux density
at 20-28 $\mu$m (circles) and 28-32 $\mu$m (blue diamonds).
The time origin of the lower-panel observations is 551 seconds later
than that of the upper.  }

\end{figure}

\begin{figure}
\center
\resizebox{\hsize}{!}{\includegraphics{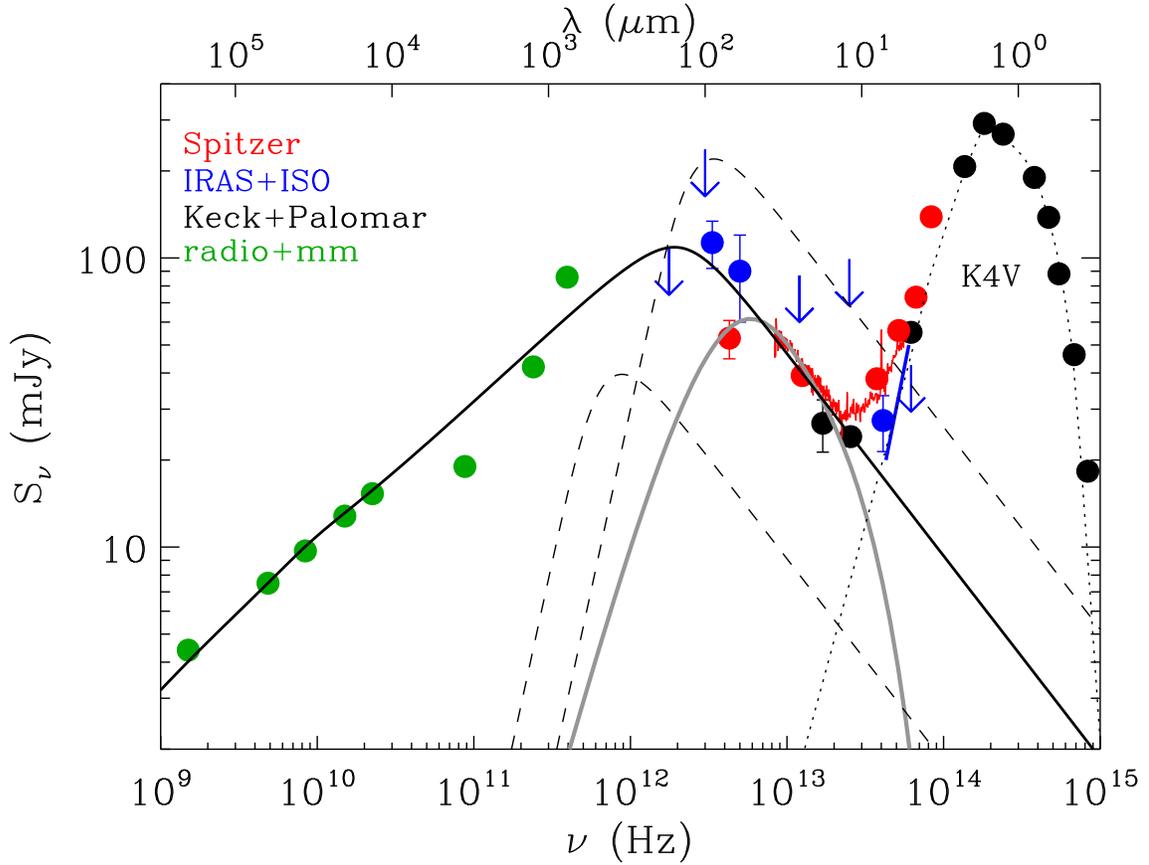}}
\caption{Overall spectral energy distribution of AE Aqr. Spitzer
measurements from this work are shown in red (large dots for IRAC
and MIPS, red line for IRS spectrum).  ISO and IRAS measurements and
upper limits are shown in blue (Abada-Simon et al. 2005). The blue line represents
the ISOCAM spectrum. The black dots show the lower values of the 
UBVRIJHK measurements obtained at Palomar and the average M, 11.6
$\mu$m and 17.6 $\mu$m measurements obtained with {\it Keck} in 2002
(Dubus et al. 2004). The black dotted line is a K4V star fit to the 
optical and near-IR data illustrated for comparison. The 
green dots are average radio and
mm fluxes taken from Abada-Simon et al. (1993, 2005). The solid black line is the
average emission from clouds ejected on a timescale $t_{\rm
flare}$=10~s and cooling adiabatically on the same timescale
(\S\ref{nonthermal}). The dashed lines show the instantaneous
spectrum emitted by a single cloud at $t=0$ and at $t=10$~s, just
before a new cloud flares. Alternatively, the solid grey line is the
emission from an optically thick circumbinary disk with $T\propto
R^{-0.54}$ and a temperature of 55~K at its maximum radius of 1.2 AU
(\S\ref{thermal}).}
\end{figure}

\end{document}